\documentclass[preprint,review,12pt,number,sort&compress]{elsarticle}

\usepackage{graphicx}
\usepackage{color}
\usepackage{ulem}

\usepackage{amssymb}
\usepackage{threeparttable}

\usepackage{url}

\journal{Astroparticle Physics}

\begin{document}

\begin{frontmatter}



\title{Surrounding material effect on measurement of thunderstorm-related neutrons}

\author[add1,add2]{H. Tsuchiya\corref{cor1}}
 \cortext[cor1]{Tel: +81-292825458, E-mail: htsuchiya@riken.jp}

\address[add1]{High-energy Astrophysics Laboratory, RIKEN Nishina Center, \\
2-1, Hirosawa, Wako, Saitama 351-0198, Japan}

\address[add2]{Nuclear Science and Engineering Directorate, Japan Atomic Energy Agency,\\
2-4 Shirakata Shirane, Tokai-mura, Naka-gun, Ibaraki 319-1195, Japan}
\begin{abstract}
Observations of strong flux of low-energy neutrons were made by $^{3}\mathrm{He}$ counters during thunderstorms [Gurevich et al (Phys. Rev. Lett. 108, 125001, 2012)].
How the unprecedented enhancements were produced remains elusive. 
To better elucidate the mechanism, a simulation study of surrounding material impacts on measurement by $^{3}\mathrm{He}$ counters was performed with GEANT4. 
It was found that unlike previously thought,  a $^3\mathrm{He}$ counter had a small sensitivity to high-energy gamma rays because of inelastic interaction with
its cathode-tube materials (Al or stainless steel). A $^{3}\mathrm{He}$ counter with the intrinsic small sensitivity, if surrounded by thick materials, would largely detect
thunderstorm-related gamma rays rather than those neutrons produced via photonuclear reaction in the atmosphere. On the other hand, the counter, 
if surrounded by thin materials and located away from a gamma-ray source,
would observe neutron signals with little gamma-ray contamination. Compared with the Gurevich measurement, 
the present work allows us to deduce that the enhancements are attributable to gamma rays, if
their observatory was very close to or inside a gamma-ray emitting region in thunderclouds.
\end{abstract}

\begin{keyword}
$\mathrm{^{3}He}$ counter \sep neutron detection  \sep gamma-ray detection \sep thunderstorms 
 \end{keyword}

\end{frontmatter}
\section{Introduction}\label{sec:intro}
Like the Sun and supernova remnants, thunderclouds as well as lightning are powerful particle accelerators in which
electrons are accelerated by electric fields to a few tens of MeV or higher energies. Then, 
they in turn produce high-energy gamma rays extending from a few hundred keV to
a few tens of MeV or 100 MeV on rare occasions. 

In addition to gamma rays and electrons, some observations~\cite{Shah_1985,armenia_2010, armenia_2012,Tibet_2012}
showed that neutrons were probably produced in association with lightning and thunderclouds.
To explain such neutron generations, 
two mechanism have been investigated theoretically and experimentally since the first positive neutron detection~\cite{Shah_1985}. 
One is fusion mechanism via $\mathrm{^2H}+\mathrm{^2H} \rightarrow \mathrm{n} + \mathrm{^{3}He}$, and 
the other is  photonuclear reaction or the Giant Resonance Reaction (GDR),  
mainly via $^{14}\mathrm{N} + \gamma (>10.6\,\, \mathrm{MeV}) \rightarrow \mathrm{n} + ^{13}\mathrm{N}$ in 
the atmosphere. Conducting numerical calculations, \citet{BR_2007_Npro} presented that only the latter was 
feasible in an usual thunderstorm environment.
However, a recent calculation considering ion runaway in a lightning discharge suggested a possibility of neutron production via  the former~\cite{IonRunaway}. 
Thus, a neutron generation process in thunderstorms remains elusive. 

Experimentally, a $\mathrm{BF_{3}}$ and $^3\mathrm{He} $ counters were frequently employed
in order to detect neutrons associated with thunderstorms. As well known, 
the two detectors have high sensitivity to neutrons thanks to high total cross-section in thermal to epithermal
energy region; 3840 b for $^{10}\mathrm{B}$ and 5330 b for $^3\mathrm{He}$ at 0.025 eV~\cite{Knoll}.  Especially, 
a set of neutron monitors (NMs), installed at high mountains with an altitude of $>$3000 m,  detected remarkable count increases during thunderstorms~\cite{armenia_2010,Tibet_2012}. Generally, 
a NM consists of a $\mathrm{BF_{3}}$ counter and its thick shields of lead and polyethylene [$\mathrm{(C_2H_4)_{n}}$]~\cite{NM64_1,NM64_2}. 
Thus, it was naturally considered that the detected count increases by NMs were attributable to neutrons, 
not gamma rays. However, \citet{Tibet_2012}, using GEANT4 simulations~\cite{GEANT4},
demonstrated that such a NM had a low but innegligible sensitivity to gamma rays with their energy higher than 7 MeV
because they can produce neutrons in the surrounding lead blocks via photonuclear reaction. 
Consequently, they pointed out that count enhancements of NMs associated with thunderstorms 
were dominated by gamma rays rather than neutrons. This claim was favored shortly afterward by \citet{armenia_2012}.

\begin{figure}
\begin{center}
\noindent\includegraphics[scale=0.5]{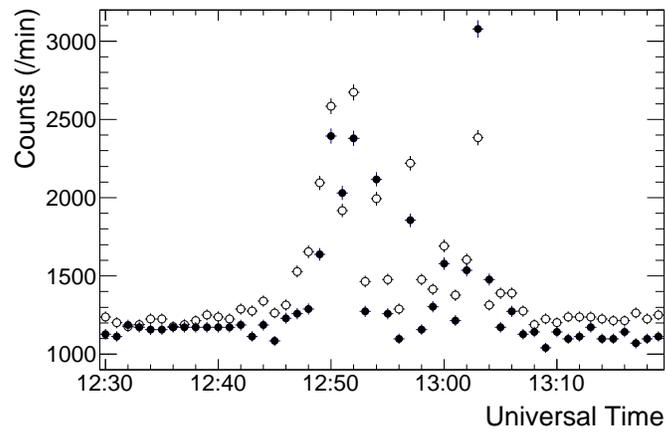}
 \caption{A thunderstorm event obtained by \citet{Gurevich_Obs2012} on 2010 August 10. Plotted data are taken from Figure 1 in the Gurevich paper.
 Open and filled circles denote one minute count histories of external and internal counters. As mentioned later, external and internal counters are
 called in this paper plywood and Fe$+$C ones, respectively. 
 }
 \label{fig:ex_event}
\end{center}
\end{figure}

As shown in Figure~\ref{fig:ex_event}, \citet{Gurevich_Obs2012} recently reported detections of strong flux of
low-energy ($<$ a few keV) neutrons during thunderstorms. 
They observed the enhancements by several independent detectors for 1 minutes or longer, in coincidence with 
high electric field changes ($<\pm 30\, \mathrm{kV/m}$).
Such a long duration, together with the simultaneous detections, may exclude the increases being due to electrical noise, and is similar to prolonged ones 
observed by other groups~\cite{armenia_2010,Tibet_2012}. Unlike the other observations, 
the Gurevich's events were done with a set of $\mathrm{^{3}He}$ counters that were 
installed at a high mountain with an altitude of 3340 m.  They argued that the detected neutron 
flux of 0.03$-$0.05 $\mathrm{cm^{-2}s^{-1}}$ were not able to be explained by the photonuclear reaction,
requiring at least three orders of magnitude higher flux of gamma rays emission than previously measured. 
However, such an increase obtained by $\mathrm{^{3}He}$ counters may originate from gamma rays, not neutrons, if we consider inelastic interaction
between high-energy gamma rays and their cathode wall made by aluminum or stainless steel. For example, 
a threshold energy of $^{27}\mathrm{Al}(\gamma, n)^{26}\mathrm{Al}$ and $^{27}\mathrm{Al}(\gamma, p)^{26}\mathrm{Mg}$
is 13.1 MeV and 8.3 MeV, respectively~\cite{IAEA_2000}. Actually, gamma rays at enrages of 10 MeV or higher have been 
measured by sea-level experiments~\cite{Dwyer_10MeV,TERA_2012,growth_2007,growth_2011,growth_2013}, 
high-mountain ones~\cite{norikura_2009,Fuji_2009,armenia_2010, armenia_2011,Tibet_2012},
and space missions~\cite{RHESSI_2009,FERMI_2010,AGILE_100MeV}. In addition, 
it is well known that neutron measurement by a $^3\mathrm{He}$ counter is disturbed by gamma rays
in a mixed field of gamma rays and neutrons~\citep{He3_gamma1,He3_gamma2}. 
Such a mixture environment is similar to observations of gamma rays and neutrons 
during thunderstorms.

In this paper, we investigate how materials surrounding $^{3}\mathrm{He}$ counters affect their measurement
during thunderstorms. 
For this aim, we derived in Section 2, with GEANT4, detection efficiency of a $^{3}\mathrm{He}$ counter for $>$10 MeV gamma rays as well as 
neutrons in 0.01 eV$-$20 MeV energy range. 
Some authors~\citep{armenia_2012,BabichCal_2013} argued against an interpretation given by \citet{Gurevich_Obs2012}, but did not clearly gave detection 
efficiency of a $^{3}\mathrm{He}$ counter for gamma rays. 
Then, to examine how neutrons and gamma rays contributes to a $^{3}\mathrm{He}$ counter surrounded by a thick or thin material, 
we utilized two roof configurations according to \citet{Gurevich_Obs2012} in Section 3. 
Considering the derived efficiency and roof effects on neutron detection during thunderstorms, 
we argue the Gurevich observations. 

\section {Detection efficiency of a $\mathrm{^{3}He}$ counter}
As described in \citep{He3_gamma2}, a reason why a $\mathrm{^{3}He}$ counter has a sensitivity to gamma rays is believed that 
they occasionally supply either neutrons or protons in the counter via inelastic interaction with a cathode wall.
As a consequence, such a gamma-ray induced nucleon would produce a large energy deposit in the counter. 
Table~\ref{tab:PNreaction} lists properties of several photonuclear reactions to be considered in this paper. 
From this table, gamma rays at energies of $>$10 MeV are found to probably give a contribution to a $^{3}\mathrm{He}$ counter
during thunderstorms, because its cathode usually consists of either Al or stainless steel. 

\begin{figure}
\begin{center}
\noindent\includegraphics[scale=0.3]{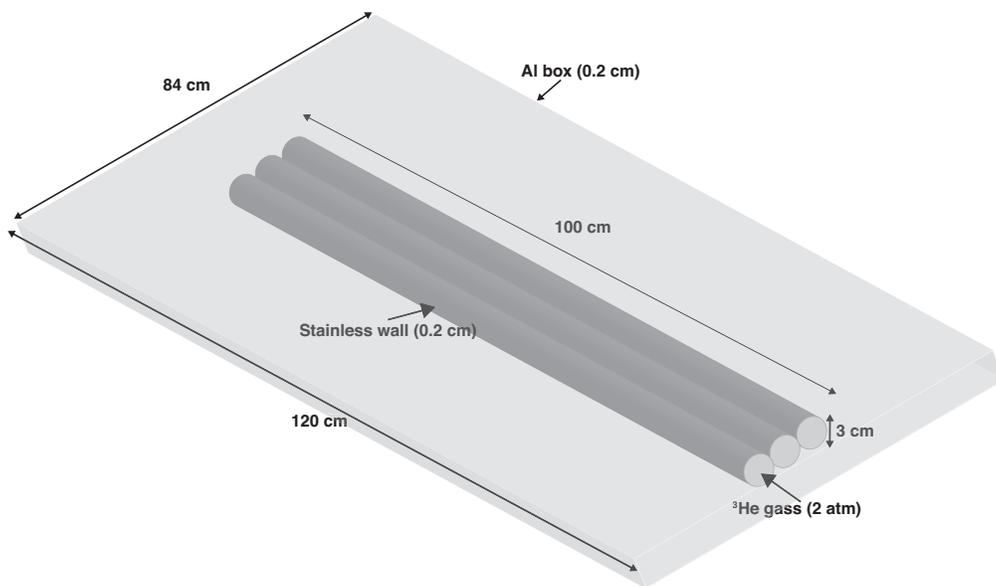}
 \caption{A schematic view of the Gurevich $^{3}\mathrm{He}$ counters. }
 \label{fig:Gurevich_counter}
\end{center}
\end{figure}

For the purpose of calculating detection efficiencies of $^{3}\mathrm{He}$ counters
for neutrons and gamma rays in the relevant energy range,   
we adopted in the GEANT4 simulation a hadronic model of QGSP\verb+_+BERT\verb+_+HP and GEANT4 standard electromagnetic physics package
to simulate neutron reactions and electromagnetic interactions including GDR, respectively.
Then, we constructed a set of three $\mathrm{^3He}$ counters confined in an Al box with an area of $1.2 \times 0.84\,\,\mathrm{m^2}$
based on "Experimental setup" of \citep{Gurevich_Obs2012} and a reference given by Gurevich group~\citep{TienShanDet}. The setup
is shown in Figure~\ref{fig:Gurevich_counter}.
Each counter has a diameter of 3 cm and a length of 100 cm, containing 100\% $\mathrm{^3He}$ gas
with a pressure of 2 atm. Because the thickness and cathode material were not shown in \citep{Gurevich_Obs2012,TienShanDet}, 
we employed in our GEANT4 simulation 2-mm thick stainless steel (74\%Fe + 8\%Ni + 18\%Cr) that is generally used by 
a commercial $\mathrm{^3He}$ counter.
Then, $10^{6}$ neutrons or $10^{7}$ gamma rays with mono energy were
 illuminated on the same area of a set of six $\mathrm{^{3}He}$ counters, isotropically injected to the counters
from the vertical to 60 degrees. 

\begin{figure}
\begin{center}
\noindent\includegraphics[scale=0.5]{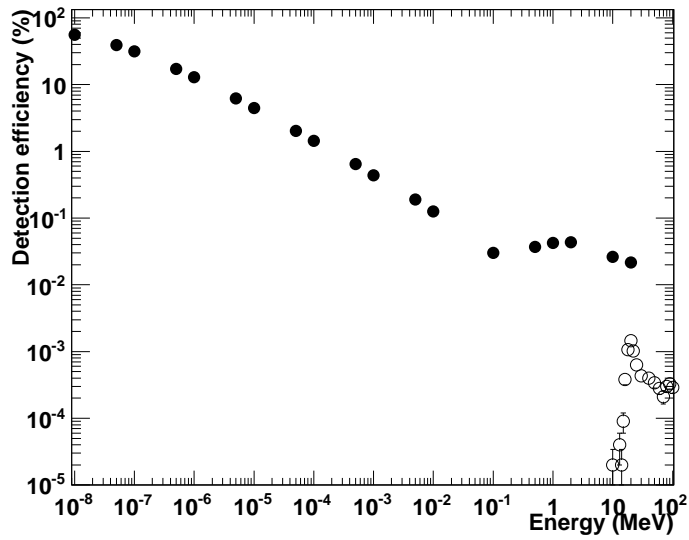}
 \caption{Detection efficiency of a $\mathrm{^{3}He}$ counter for neutrons (filled circles) and gamma rays (open circles), calculated by GEANT4. 
 The efficiencies were computed by dividing the number of events that energy deposit in a $\mathrm{^{3}He}$ counter exceeded $>$100 keV
 by total number of incident neutrons or gamma rays. 
 Quoted errors are statistical 1$\sigma$. }
 \label{fig:He3_det_ng}
\end{center}
\end{figure}

According to \citet{Gurevich_Obs2012}, an efficiency of thier $\mathrm{^3He}$ counters for neutrons in a low energy range is about 60\%, 
and the efficiency at $\sim$10 keV becomes three orders of magnitude lower. As shown in Figure~\ref{fig:He3_det_ng},
this trend is found to be consistent with that of neutron detection efficiency derived here. 
In addition, it is found that the whole structure of the detection efficiency for neutrons in a wide energy range of 0.01 eV$-$20 MeV 
completely follows the total cross-section of $^{3}\mathrm{He}$ atom\footnote{The total cross-section can be seen at e.g. \url{http://wwwndc.jaea.go.jp/j40fig/jpeg/he003_f1.jpg}};
mainly a neutron capture reaction of $^{3}\mathrm{He(n, p)T}$ in energy below 0.1 MeV and an elastic scattering above 0.1 MeV.
These consistencies validate the simulation.

Due to the smaller cross-section,  gamma rays are detected with a relatively low sensitivity of at most $(1.47\pm0.12)\times10^{-3}\%$ at 20 MeV (the error is statistical one only).
This is consistent with that each peak energy of photonuclear reaction for $^{52}\mathrm{Cr}$ and $^{56}\mathrm{Fe}$ is around 20 MeV (Table~\ref{tab:PNreaction}).
From this simulation, it was found that gamma-ray induced protons or neutrons (alpha on rare occasions) had a typical kinetic energy of nearly 10 MeV. 
Then, such a proton (or alpha) deposits via ionization loss an amount of a hundred keV or higher energies in a $^{3}\mathrm{He}$ counter, while the gamma-ray induced
neutron mainly causes an elastic scattering with $^{3}\mathrm{He}$ nucleus to produce a large energy deposit of $>$1 MeV. 
Changing a cathode material of stainless steel to Al, we found that gamma-ray detection efficiency for Al was the same with the derived
values (Fig.~\ref{fig:He3_det_ng}), within statistical uncertainty. 

\section{Results}
In this paper, we utilized secondary energy spectra obtained in the same manner with \citet{Tibet_2012}. 
Air density of their observational level (4300 m above sea level) is $\sim$$8\times10^{-4}$ $\mathrm{g\,cm^{-3}}$,
which is almost the same with that of $\sim$$9\times10^{-4}$ $\mathrm{g\,cm^{-3}}$ (3340 m) of \citet{Gurevich_Obs2012}. 
Based on results of actual measurement~\citep{AGILE_100MeV, growth_2011}, 
these spectra were made with GEANT4 assuming that primary gamma rays have power-law type energy 
spectrum with its index, $\beta$ of $-1$, $-2$, or $-3$ and the energy range of 10$-$100 MeV. 
Gamma rays arriving at the observational level
have energy spectra that are almost the same with those of Fig.7 in \citet{Tibet_2012}. Spectra of neutrons will be shown later. 

\subsection{Survival probability of neutrons}
From the viewpoint that neutron production is a well-known photonuclear reaction, we
consider two possible reasons of the enhancements presented by \citet{Gurevich_Obs2012}.
One is that low-energy neutron flux produced via photonuclear reaction is very high, 
especially in thermal to epithremal energies (0.01$-$10 eV), in which a $^{3}\mathrm{He}$ counter readily detect neutrons. 
The other is a gamma-ray contribution in detected counts.
Here, we firstly investigated the former case by deriving survival probability of $>$0.01 eV neutrons at the observational level.

\begin{figure}
\begin{center}
\noindent\includegraphics[scale=0.5]{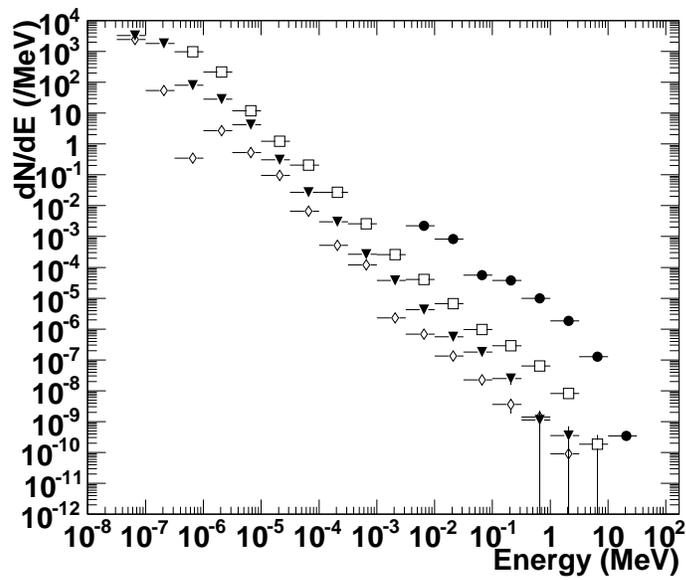}
 \caption{Energy spectra of neutrons reaching the observational level assuming $\beta=-2$ and $H$ = 0.01 (circles),
  0.3 (squares), 1.5 (triangles), and 3 (diamonds) km.}
 \label{fig:TopOfRoof_Nspe}
\end{center}
\end{figure}

\begin{figure}
\begin{center}
\noindent\includegraphics[scale=0.5]{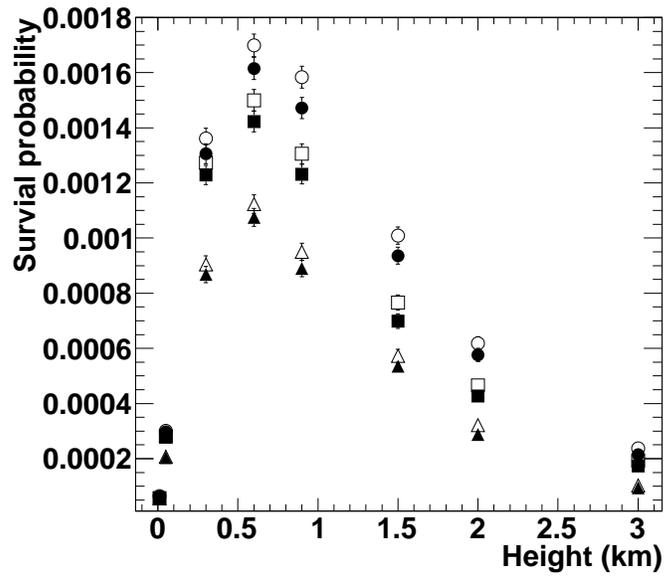}
 \caption{Comparison between survival probability of $>$0.01 eV neutrons (open symbols) arriving at the observational level and that for $>$1 keV neutrons (filled symbols).
 Circles, squares, and triangles represents $\beta$ of $-1$, $-2$, and $-3$, respectively. 
 The horizontal axis shows assumed source height in km. Error bars attached to individual data points are statistical 1$\sigma$. }
 \label{fig:comp_surb}
\end{center}
\end{figure}

Figure~\ref{fig:TopOfRoof_Nspe} shows GEANT4-derived neutron energy spectra at the observational level,
suggesting that  path length of $<$0.3 km from a source to an observatory
is not long enough to 
thremalize neutrons produced via photonuclear reaction, 
because of the considerably long interaction length, $\sim$1.7 $\times10^{3}$ $\mathrm{g\, cm^{-2}}$ of the photonuclear reaction in the air. 
Then,
integrating each neutron energy spectrum at the observational level over 0.01 eV$-$100 MeV assuming primary gamma rays are emitted from $H$ 0.01$-$3 km,
we obtained in Figure~\ref{fig:comp_surb} survival probability of $>$0.01 eV neutrons produced via photonuclear reaction.
For comparison, survival probability of $>$1 keV neutrons are also plotted. 
As easily seen, the survival probability of $>$0.01 eV neutrons (open symbols) above 0.3 km 
is only at most 8\% higher than that for $>$1 keV (filled symbols). As expected, the $>$0.01-eV probability 
at $H<$0.3 km is almost the same with that for $>$1 keV neutrons.
Thus, we can reject a hypothesis that extremely high flux of low-energy neutrons arrived at the observational level to provide the count enhancements of \citet{Gurevich_Obs2012}.

\subsection{Comparison between flux of neutrons and gamma rays}
\begin{figure}
\begin{center}
\noindent\includegraphics[scale=0.5]{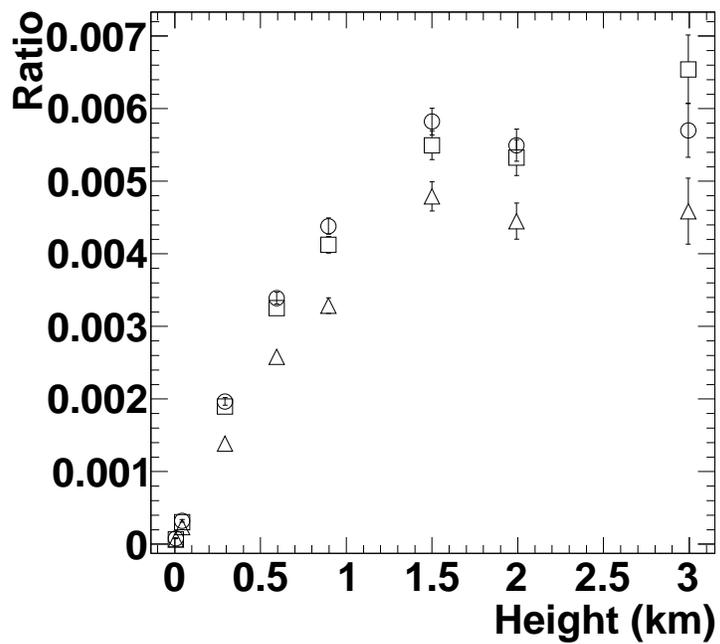}
 \caption{Ratios of flux of $>$0.01 eV neutrons at the observational level to that of $>$10 MeV gamma rays, plotted
 against source height in km. Circles, squares, and triangles represents $\beta$ of $-1$, $-2$, and $-3$, respectively.
 Errors show statistical 1$\sigma$.
 }
 \label{fig:RatioNG_H}
\end{center}
\end{figure}

We next compare flux of $>$0.01 eV neutrons reaching the observational level with that of $>$10 MeV gamma rays. 
As shown in Figure~\ref{fig:RatioNG_H}, 
the derived ratios monotonously increase from $\sim$$10^{-4}$ at $H=0.01$ km toward $\sim$5$\times10^{-3}$ at $H$ 1.5 km, and take 
roughly constant in $H$ of 1.5$-$3 km, within large statistical uncertainty.

Assuming the detected signals by $^{3}\mathrm{He}$ counters all were attributable to neutrons,
\citet{Gurevich_Obs2012} emphasized that 10$-$30 MeV gamma-ray flux of 10$-$30 $\mathrm{cm^{-2}s^{-1}}$ was required to explain
their detections of low-energy neutrons with flux of 0.03$-$0.05 $\mathrm{cm^{-2}s^{-1}}$. From their flux, we can obtain 
a ratio of neutron flux to gamma-ray one as $($1$-$5$)$$\times10^{-3}$, which is consistent with that derived above. Therefore, we may conclude that 
the relation between flux of gamma rays and neutrons follows the standpoint of photonuclear reaction, though
we do not still know how such high gamma-ray flux is generated in thunderclouds.

\subsection{Contribution ratio}
\subsubsection{Configurations of roof and rooms}
For the purpose of calculating contribution ratios of neutrons and gamma rays in $^{3}\mathrm{He}$-counter signals
during thunderstorms, hereafter we employed energy spectra of secondary neutrons and gamma rays assuming
$\beta=-2$.  

In the measurements done by \citet{Gurevich_Obs2012}, one group of $^3\mathrm{He}$ counters was installed inside a building with its roof
consisting of iron and carbon (This group is called "internal counters" in \citep{Gurevich_Obs2012}). The thickness of iron and carbon is 0.2 cm and 20 cm, respectively. 
The other was arranged at a building with its roof comprised of a light plywood ($\mathrm{C_{6}H_{10}O_{5}}$) (This is called "external counters" in \citep{Gurevich_Obs2012}) . 
Because the thickness of the plywood
was not written in the literature, we assumed it was 2 cm.  Here, the area of each roof was assumed to be 10 m$\times$10 m (an area of the roofs
do not affect the final result).
Also, we adopted density of iron, carbon, and plywood as 7.87, 2.26, and 0.55 $\mathrm{g\, cm^{-3}}$, respectively. 
Then, gamma rays (neutrons) with a GEANT4-calculated energy spectrum 
were vertically irradiated from just above the individual roofs within an area of the roofs to investigate how it affects an energy spectrum of penetrating particles
including gamma rays and neutrons. Total number of gamma rays (neutrons) incident to each roof was 1$\times10^7$.

\subsubsection{Roof effect}
\begin{figure}
\begin{center}
\noindent\includegraphics[scale=0.45]{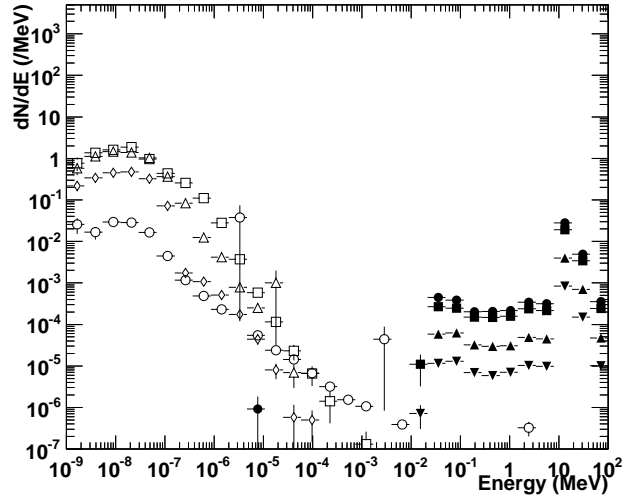}
\noindent\includegraphics[scale=0.45]{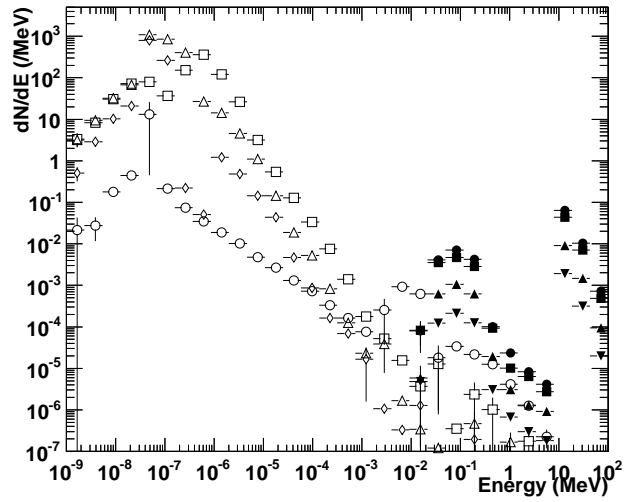}
 \caption{Energy spectra of neutrons (open symbols) and gamma rays (filled symbols) under the Fe$+$C roof (top)
 and plywood one (bottom). 
 Circles, squares, triangles, and diamonds correspond to $H$ = 0.01, 0.3, 1.5, and 3 km, respectively. 
 Errors are statistical $1\sigma$.
 }
 \label{fig:GNspe_Under_roof}
\end{center}
\end{figure}

Figure~\ref{fig:GNspe_Under_roof} shows energy spectra of neutrons and gamma rays that are observed under the individual roofs. 
Table~\ref{tab:Prob_pene_FC} and \ref{tab:Prob_pene_PL} list probability that gamma rays or neutrons incident to the two roofs penetrate through 
them. Compared with the roof-top neutron spectra (Fig.~\ref{fig:TopOfRoof_Nspe}), those under the roofs drastically decrease 
over 0.01$-$10 eV, in which neutrons can be easily detected by a $^{3}\mathrm{He}$ counter. 
The reduction is mainly because of neutron reflection by materials in the roofs. Importantly, $>$10 MeV gamma-ray flux for the Fe$+$C roof
is different from that for the plywood roof, only by a factor of $\sim$2.

From Table \ref{tab:Prob_pene_FC} and \ref{tab:Prob_pene_PL}, we can know what kind of generation processes contribute to the spectra under 
the roofs (Fig.~\ref{fig:GNspe_Under_roof}). 
As expected, probability of gamma rays incident to each roof,  $P_\mathrm{\gamma\gamma}$ is the highest among the others  
(Table~\ref{tab:Prob_pene_FC} and \ref{tab:Prob_pene_PL}). This probability, which corresponds to $>$10 MeV power-law part 
 (filled symbols of Fig.~\ref{fig:GNspe_Under_roof}), 
indicates that part of the incident gamma rays transmit the roofs
without absorption or pair creation. $P_{\gamma\gamma}$ for Fe$+$C roof is more than 
five orders of magnitude higher than the other probabilities, and that for plywood
is more than two orders of magnitude higher than them. These results suggest that high-energy gamma rays are the most 
dominant component to enter $^{3}\mathrm{He}$ counters and can realize high gamma-ray field as previously mentioned.
Thus, a careful discrimination of thunderstorm-related neutrons from gamma rays would be required to detect neutrons
during thunderstorms, as previously pointed out by \citet{armenia_2012} as well.

In addition to $P_{\gamma\gamma}$, 
another gamma-ray component of $P_{\mathrm{n}\gamma}$ is almost related to neutron capture reaction 
of $^{14}\mathrm{N}(\mathrm{n}, \gamma)^{15}\mathrm{N}$. The reaction produces 10.8 MeV prompt gamma rays, while
other capture reactions with nuclei in the roofs do not produce $>$10 MeV prompt gamma rays, at most $\sim$8 MeV~\citep{capgam}.
In practice, it was found that neutrons captured by $^{14}\mathrm{N}$ 
were caused by elastic scatterings from the roofs.
One neutron component $P_{\gamma\mathrm{n}}$ is associated with photoinelastic reactions that produce not only neutrons
but simultaneously gamma rays at energies from a few hundred keV to a few MeV.
In addition, $P_{\mathrm{n}\mathrm{n}}$ results from incident neutrons that suffer single or multiple elastic scatterings in the roofs. Thus, 
$P_{\mathrm{n}\mathrm{n}}$ for the thin plywood roof (2 cm) is higher than that for the thick Fe$+$ C roof (20.2 cm).

\subsubsection{Detection of neutrons and gamma rays}
Convoluting the energy spectra under the individual roofs (Fig.~\ref{fig:GNspe_Under_roof}) with detection 
efficiency of a $^{3}\mathrm{He}$ counter (Fig.~\ref{fig:He3_det_ng}), we can estimate how 
neutrons and gamma rays contribute to count increases measured by a $^{3}\mathrm{He}$ counter. 

\begin{figure}
\begin{center}
\noindent\includegraphics[scale=0.5]{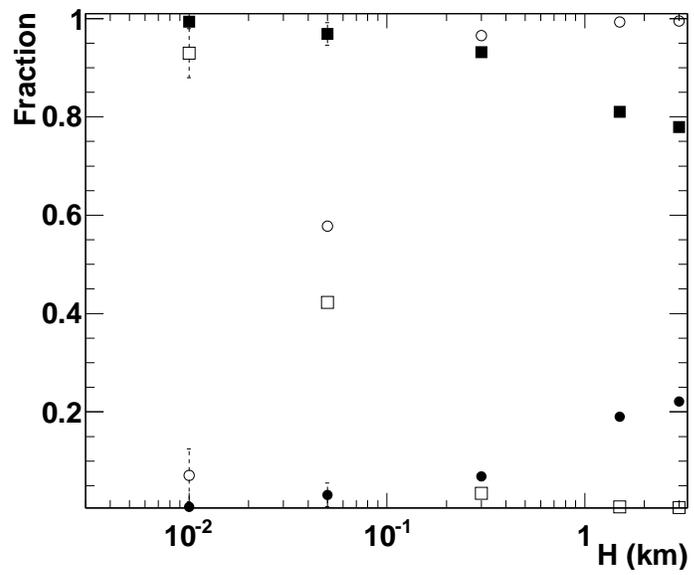}
 \caption{Contribution fraction by neutrons (circles) and gamma rays (squares) for a $^3\mathrm{He}$-coutner signal.
 Filled and open symbols correspond to the Fe$+$C and plywood roofs, respectively. Statistical 1$\sigma$ errors are attached to
 individual points. 
 }
 \label{fig:DetCompGN.eps}
\end{center}
\end{figure}

As shown in Figure~\ref{fig:DetCompGN.eps}, we found a clear difference in contribution fraction of $^{3}\mathrm{He}$-counter signals
between the two roofs. Gamma rays dominate, by $\sim$80\% or higher fraction,
signal detected by a $^3\mathrm{He}$ coutner installed under the Fe$+$C roof (filled squares) and they
contribute a tiny fraction, $<$4\% of signal for the thin plywood roof (open squares) $H$ above 0.3 km.
However, the gamma-ray fraction for the plywood roof is found rapidly increase below 0.3 km
and be more than 10 times higher at $H=0.01$ km than that for neutrons.

These results imply two important points. One is that a $^{3}\mathrm{He}$ counter covered by a thick material would have a better sensitivity to thundercloud-related
gamma rays rather than those neutrons. This is mainly because neutrons, if produced via photonuclear reaction, 
are reflected to the atmosphere by a thick materials. 
The other is that a $^{3}\mathrm{He}$ counter,
when surrounded by thin materials, largely detects thundercloud-related gamma rays if a gamma-ray source is very close to it. 

\subsection{Comparison with the Gurevich measurement}
Considering both the neutron and gamma-ray contributions, we can calculate a ratio of detected counts under the plywood roof, $N_\mathrm{out}$,
to those for the Fe$+$C roof, $N_\mathrm{in}$ and compare them with the measurement done by \citet{Gurevich_Obs2012}.
Obviously, either $N_\mathrm{in}$ or $N_\mathrm{out}$ is shown by 
\begin{equation}
N_k = N_{\mathrm{n},k} + N_{\mathrm{\gamma}, k}, \,\,\, k = \mathrm{in}, \mathrm{out}, 
\end{equation}
where $N_{\mathrm{n},k}$ and $N_{\mathrm{\gamma}, k}$ represent counts expected from neutrons and gamma rays, respectively. 
Then, $N_{\mathrm{n}, k}$ and $N_{\mathrm{\gamma}, k}$ is expressed by 
\begin{eqnarray}
N_{\mathrm{n}, k} &=& \alpha A \int_\mathrm{0.001\, eV} ^\mathrm{100\, MeV} \eta_\mathrm{n}(E_\mathrm{n})\epsilon_\mathrm{n}(E_\mathrm{n})dE_\mathrm{n} \\
N_{\mathrm{\gamma}, k} &=& \alpha A \int_\mathrm{10\, MeV} ^\mathrm{100\, MeV} \eta_\mathrm{\gamma}(E_\mathrm{\gamma})\epsilon_\mathrm{\gamma}(E_\mathrm{\gamma})dE_\mathrm{\gamma},
\end{eqnarray}
respectively. Here, $\alpha$ and $A$ shows a normalization constant of primary gamma-ray spectrum at source and geometrical area of $^{3}\mathrm{He}$ counters, respectively. 
$\eta$ and $\epsilon$ are an energy spectrum under each roof (Fig.~\ref{fig:GNspe_Under_roof}) and detection efficiency of a $^{3}\mathrm{He}$ counter for neutrons and gamma rays (Fig.\ref{fig:He3_det_ng}), respectively.
Finally, we computed ratios of $R_\mathrm{n}$, $R_\mathrm{\gamma}$, and $R_\mathrm{T}$ as 
$N_\mathrm{n,out}/N_\mathrm{n,in}$, $N_\mathrm{\gamma,out}/N_\mathrm{\gamma,in}$, and $N_\mathrm{out}/N_\mathrm{in}$ to eliminate unknown $\alpha$.

\begin{figure}
\begin{center}
\noindent\includegraphics[scale=0.5]{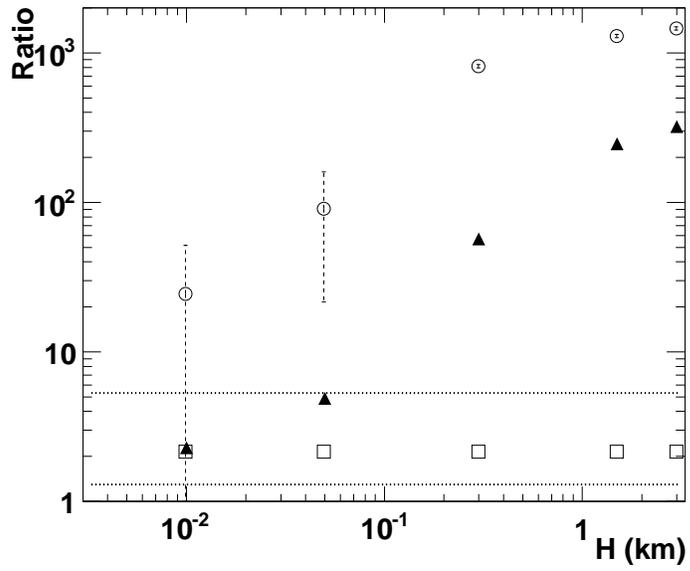}
 \caption{Ratios of a count expected under the plywood roof to that for the Fe$+$C one, plotted against H (km).
  Circles, squares, and triangles show $R_\mathrm{n}$, $R_\mathrm{\gamma}$, and $R_\mathrm{T}$ (see text).
  Area between two horizontal dotted lines denotes the range of ratios determined by measurement of \citet{Gurevich_Obs2012}. Statistical errors
 represent 1$\sigma$.
 }
 \label{fig:Ratio_GNTOT}
\end{center}
\end{figure}

Figure~\ref{fig:Ratio_GNTOT} compares calculated ratios with the measurement presented by \citet{Gurevich_Obs2012}.
Apparently, $R_\mathrm{n}$ (circles) can not reproduce the measured ratios ranging from 1.3 to 5.3 (arrow),
in a wide range of $H$ (0.01$-$3 km).
\citet{armenia_2012} also remarked that they were unable to explain the Gurevich results when considering a neutron contribution originating 
from photonuclear reaction. 

Very interestingly,  $R_\mathrm{\gamma}$, which is almost constant in a wide range of $H$, is quite consistent with the Gurevich results. 
Consequently, $R_\mathrm{T}$ matches the Gurevich results when $H$ is less than 0.05 km.
From this as well as the gamma-ray contribution at $H<$ 0.05 km (Fig.~\ref{fig:DetCompGN.eps}), 
we guess that the observatory (3340 m above sea level) of \citet{Gurevich_Obs2012} was very close to or inside a gamma-ray source region in thunderclouds
and hence their $^{3}\mathrm{He}$ counters almost detected thundercloud-related gamma rays.
In such a nearby case, a traveling path for gamma rays to reach their counters would be too short to produce sufficient neutron flux via photonuclear 
reaction in comparison with gamma-ray flux (squares in Fig.~\ref{fig:RatioNG_H}). 
In addition, \citet{norikura_2009}, observing thundercloud-related gamma rays at an observatory located at 2770 m above sea level,  
demonstrated that a source height of the gamma-ray emitting region was 0.06$-$0.13 km at 95\% confidence level. 
Therefore,  an extremely nearby situation would be expected in case of a mountain observatory.

\section{Conclusions}
The present simulation clearly showed that a $^{3}\mathrm{He}$ counter had a small sensitivity to $>$10 MeV gamma rays.
It was found that this ability enabled $^{3}\mathrm{He}$ counters to detect thundercloud-related gamma rays rather than those neutrons if surrounded by thick materials.
Thus, it would be rather difficult to conclude that a $^{3}\mathrm{He}$-counter signal detected during thunderstorms
is all attributable to neutrons like previously thought~\citep{Gurevich_Obs2012}. To obtain a conclusive answer whether detected counts are dominated by neutrons or gamma rays, 
we must consider a source height as well as surrounding material impacts on measurement by $^{3}\mathrm{He}$ counters.
Given the present results,  we may conclude that the large count enhancements obtained by \citet{Gurevich_Obs2012} is resulted from  
$>$10 MeV gamma rays radiated from a very nearby source in thunderclouds. 

To clarify the present finding, we will need to install $^{3}\mathrm{He}$ counters at other high mountains 
and conduct further experiments with $^{3}\mathrm{He}$ 
counters and gamma-ray detectors. In addition, like a recent measurement done by \citet{LabNeutron},
a laboratory experiment using a high-voltage generator and various detectors to catch neutrons and gamma rays would be promising.
In this paper, we did not consider another neutron generation process such as the fission mechanism
of $\mathrm{^2H}+\mathrm{^2H} \rightarrow \mathrm{n} + \mathrm{^{3}He}$. Therefore, we are unable to rule out a possibility that such a mechanism 
contributes to the large count enhancements. 

\section{Acknowlegements}
The present work is supported in part by JSPS KAKENHI Grant Number
24740183 (Grant-in-Aid for Young Scientists B).







%

 \clearpage
\begin{threeparttable}
\caption{Characteristics of several photonuclear reactions to be considered}
\label{tab:PNreaction}
\begin{tabular}{ccccc}\hline
Nuclide\tnote{1}     &     $E_{\mathrm{n}}$ (MeV)\tnote{2}  &  $E_{\mathrm{p}}$ (MeV)\tnote{3} &  $E_{\mathrm{peak}}$ (MeV)\tnote{4}&$\sigma_{\mathrm{peak}}$(mb)\tnote{5}   \\
$^{12}\mathrm{C}$   &      18.7                                &    16.0                                             & 23                               & 20 \\
$^{14}\mathrm{N}$   &      10.6                                &      7.6                                             & 23                                &  27 \\
$^{16}\mathrm{O}$   &      15.7                               &     12.1                                            &  22                               &   31 \\
$^{27}\mathrm{Al} $  &   13.1                               &        8.3                                           &  21                                & 42 \\
$^{52}\mathrm{Cr}$  &   12.0                               &         10.5                                          &  20                               & 95 \\
$^{56}\mathrm{Fe}$ &     11.2                              &         10.2                                         &  20                                & 80 \\ \hline
\end{tabular} 
\begin{tablenotes}
\item[1] These values were gathered from \citep{IAEA_2000}.
\item[2] Threshold energy of ($\gamma$, n) reaction.
\item[3] Threshold energy of ($\gamma$, p) reaction.
\item[4]  Peak energy of total photonuclear reaction.
\item[5]  Cross-section at peak energy. 
\end{tablenotes}
\end{threeparttable}

\begin{threeparttable}
\caption{Probability that gamma rays ($>$10 MeV) and neutrons ($>$0.001 eV) penetrate under the Fe$+$C roof. }
\label{tab:Prob_pene_FC}
\begin{tabular}{ccccc}\hline
$H$\tnote{a}          &  $P_{\gamma\gamma}\tnote{b}$ & $P_{\gamma\mathrm{n}}$\tnote{c} & $P_{\mathrm{n}\gamma}$\tnote{d} &$P_{\mathrm{n}\mathrm{n}}$\tnote{e} \\ 
0.01 &   0.43\tnote{f} & $(1.1\pm0.3)\times10^{-6}$&  $(4.33\pm0.14)\times10^{-9}$ &$(7.44 \pm 0.07)\times10^{-8}$ \\
0.3   &      0.30\tnote{f}    &    $(5.4\pm1.9)\times10^{-7}$    & $(4.85\pm0.07)\times10^{-7}$ & $ (2.82 \pm 0.06)\times10^{-7}$\\
1.5   &      0.062\tnote{f}  &    $(1.3\pm 0.4)\times10^{-7}$  & $(4.38\pm0.05)\times10^{-7}$   & $(1.43\ \pm 0.03)\times10^{-7}$\\
3      &      0.013\tnote{f}  &    $(9 \pm 5)\times10^{-9}$       & $(1.33\pm0.01)\times10^{-7}$   & $(3.38 \pm 0.08)\times10^{-8}$\\ \hline
\end{tabular} 
\begin{tablenotes}
\item[a] Assumed source height in km.
\item[b] Probability that gamma rays penetrate through the roof in case of gamma-ray incidence.
\item[c] Probability that neutrons penetrate through the roof in case of gamma-ray incidence.
\item[d] Probability that gamma rays penetrate through the roof in case of neutron incidence.
\item[e] Probability that neutrons penetrate through the roof in case of neutron incidence.
\item[f] Not shown, but the statistical error is less than 0.1\%.
\item[*] All quoted errors are due only to Monte Carlo statistics.
\end{tablenotes}
\end{threeparttable}
\clearpage 

\begin{threeparttable}
\caption{Probability that gamma rays ($>$10 MeV) and neutrons ($>$0.001 eV) penetrate under the plywood roof.}
\label{tab:Prob_pene_PL}
\begin{tabular}{ccccccccc}\hline
$H$\tnote{a}          &  $P_{\gamma\gamma}\tnote{b}$ & $P_{\gamma\mathrm{n}}$\tnote{c} & $P_{\mathrm{n}\gamma}$\tnote{d} &$P_{\mathrm{n}\mathrm{n}}$\tnote{e} \\ 
0.01 &    0.96\tnote{f} & $(3.9\pm1.4)\times10^{-6}$&  $(2.20\pm0.22)\times10^{-9}$ & $(2.781 \pm 0.003)\times10^{-5}$ \\
0.3 & 0.66\tnote{f}                      &  $(1.3\pm0.7)\times10^{-6}$ & $(3.17\pm0.13)\times10^{-7}$  & $(4.618\pm0.005)\times10^{-4}$ \\
1.5 &0.14\tnote{f}                       &  $(3.5\pm1.6)\times10^{-7}$  & $(2.45\pm0.09)\times10^{-7}$  &  $(2.569\pm0.003)\times10^{-4}$\\
3    & 0.029\tnote{f}                    &  $(1.3\pm0.4)\times10^{-7}$  & $(6.9\pm0.2)\times10^{-8}$ & $(5.954\pm0.008)\times10^{-5}$ \\ \hline
\end{tabular} 
\begin{tablenotes}
\item[a] Assumed source height in km.
\item[b] Probability that gamma rays penetrate through the roof in case of gamma-ray incidence.
\item[c] Probability that neutrons penetrate through the roof in case of gamma-ray incidence.
\item[d] Probability that gamma rays penetrate through the roof in case of neutron incidence.
\item[e] Probability that neutrons penetrate through the roof in case of neutron incidence.
\item[f] Not shown, but the statistical error is less than 0.1\%.
\item[*] All quoted errors are due only to Monte Carlo statistics
\end{tablenotes}
\end{threeparttable}
\clearpage 
\clearpage 


\end{document}